\documentclass[a4paper,11pt,fleqn]{article}

\usepackage{amsmath,amssymb,amsthm,epsfig,graphics} 

\newcommand{\todo}[1][\null]{\ensuremath{\clubsuit}}

\newcommand{\noprint}[1]{}

\newcommand{\D}{\,{\rm d}}
\DeclareMathOperator{\erf}{erf}
\newcommand{\const}{\mathop{\rm const}\nolimits}

\newcommand{\sgn}{\mathop{\rm sgn}\nolimits}

\newcommand{\lsemioplus}{\mathbin{\mbox{$\lefteqn{\hspace{.77ex}\rule{.4pt}{1.2ex}}{\in}$}}}

\newcounter{tbn}

\newcounter{mcasenum}

\newtheorem{theorem}{Theorem}

\newtheorem{corollary}{Corollary}

\newtheorem*{proposition*}{Proposition}
{\theoremstyle{definition}

\newtheorem{remark}{Remark}

}

\begin{document}
\par\noindent {\LARGE\bf Group classification and exact solutions of
\\
variable-coefficient generalized Burgers equations
\\
with linear damping\par}

\vspace{6mm}

{\par\noindent\large Oleksandr A.
Pocheketa$^{\dag}$, Roman O.
Popovych$^{\dag\ddag}$ and Olena O.~Vaneeva$^{\dag}$ \par\vspace{2mm}\par}

\vspace{2mm}

{\par\noindent\it ${}^\dag$\;Institute of Mathematics of National Academy of Sciences of Ukraine,
\\
$\phantom{{}^\dag}$\;3 Tereshchenkivska Str., Kyiv-4, 01601 Ukraine
\\[2ex]
${}^\ddag$\;Wolfgang Pauli Institut, Universit\"at Wien, Nordbergstra{\ss}e 15,
A-1090 Wien, Austria }

{\vspace{4mm}\par\noindent
$\phantom{{}^\dag{}}$E-mail: \it pocheketa@yandex.ua, rop@imath.kiev.ua, vaneeva@imath.kiev.ua \par}

{\vspace{7mm}\par\noindent\hspace*{5mm}\parbox{150mm}{\small
Admissible point transformations between Burgers equations with linear damping and time-dependent coefficients
are described and used in order to exhaustively classify Lie symmetries of these equations.
Optimal systems of one- and two-dimensional subalgebras of the Lie invariance algebras obtained are constructed.
The corresponding Lie reductions to ODEs and to algebraic equations are carried out.
Exact solutions to particular equations are found.
Some generalized Burgers equations are linearized to the heat equation by composing equivalence transformations with the Hopf--Cole transformation.
}\par}

\vspace{3mm}

\section{Introduction}

The remarkable equation $u_t+uu_x=u_{xx}$
was suggested by Burgers in 1948~\cite{burg1948a} as a model for one-dimensional turbulence.
The equation bears his name now, though it appeared in works by Forsyth~\cite{fors1959a} and Batemen~\cite{bate1915a} earlier.
Various generalizations of the Burgers equation are used as models in nonlinear acoustics.
A number of such models were reviewed in~\cite{sach2009a}; in particular, see p.~40 for the discussion on  appearance of a linear damping term.
The equation that includes effects of cylindrical or spherical spreading and of non-equilibrium relaxation,
\begin{gather*}
u_t+uu_x+\frac{j}{2t} u=\frac{\delta}2u_{xx},
\quad
\delta=\const,
\end{gather*}
was proposed in~\cite[Chapter~4]{leib1974a} and studied later in~\cite{crig1979a,sach1994a}.
Here the integer $j$ is the number of dimensions the wave can spread
($j=1$ for cylindrical spreading and $j=2$ for spherical one).

The generalized Burgers equation
describing the propagation of weakly nonlinear acoustic waves under the impact
of geometrical spreading and thermoviscous diffusion was derived in~\cite{hamm1989a}.
In dimensionless variables it can be reduced to the form
\begin{gather*}
u_t+uu_x=g(t)u_{xx},\quad
g\ne0.
\end{gather*}

As Lie symmetries provide a powerful tool for finding exact solutions, such model equations were
intensively studied from the Lie symmetry point of view.
For example, Lie symmetries of the latter Burgers equations with variable diffusion coefficient
were considered in~\cite{doyl1990a} and~\cite{wafo2004d}.
The complete group classification of the class
\begin{gather*}
u_t+a\big(u^m\big)_x=g(t)u_{xx},
\quad
ag\ne0,
\quad
m\ne0,1,
\end{gather*}
was recently carried out in~\cite{vane2013b}.
Self-similar solutions of the equations
\begin{gather*}
u_t+u^nu_x+\frac{j}{2t} u=\frac{\delta}2u_{xx}, \quad n>0,\quad j\geqslant0,\quad \delta>0,
\end{gather*}
were investigated in~\cite{rao2002a}.
The paper~\cite{vaga2006a} concerns
Lie symmetries and Lie reductions of generalized Burgers equations with linear damping and variable viscosity,
\begin{gather*}
u_t+u^nu_x+\alpha u=g(t)u_{xx},
\quad
g\ne0,
\quad
n\in\mathbb{R},
\quad
\alpha\geqslant0.
\end{gather*}

In the present paper we investigate Lie symmetries and construct exact solutions of
variable coefficient generalized Burgers equations with linear damping of the form
\begin{gather}\label{GBE}
u_t+u^nu_x+h(t)u=g(t)u_{xx}, \quad ng\ne0.
\end{gather}
Here $h(t)$ and $g(t)$ are arbitrary smooth functions with $g\neq0$, and $n$ is an arbitrary nonzero constant.
This class includes all aforementioned equations as special cases.

Note that the more general class of variable coefficient generalized Burgers equations with three time-dependent coefficients of the form
\begin{gather}\label{GBEfgh}
u_t+f(t)u^nu_x+h(t)u=g(t)u_{xx}, \quad nfg\ne0,
\end{gather}
reduces to the class~\eqref{GBE} via a change of the variable $t$.
Therefore, without loss of generality it is sufficient to study  the class~\eqref{GBE},
since all results on symmetries, exact solutions, conservation laws, etc.
for the class~\eqref{GBEfgh} can be derived from those
obtained for the class~\eqref{GBE}.

Our aim is to present the complete classification of Lie symmetries of the class~\eqref{GBE},
to enhance existing results on Lie symmetries of its certain subclasses
and to find, when possible, exact solutions to particular equations of the form~\eqref{GBE}.
The linear case $n=0$ is excluded from the consideration as exhaustively studied
from the Lie symmetry point of view~\cite{lie1881a}.
(Moreover, any linear equation of this form is reduced by a point transformation to the classical heat equation.)
We also describe equations from the class~\eqref{GBE} that are linearized to the heat equation
using the composition of equivalence transformations with the Hopf--Cole transformation.

Recent studies of group classification problems for classes of differential equations (DEs) show
that it is important to investigate the whole set of admissible transformations in a class of DEs,
which naturally possesses the structure of groupoid and is hence called the \emph{equivalence groupoid} of the class.
As a rule, the use of admissible transformations appears to be the optimal way
to get the complete group classification of a class of DEs with reasonable efforts.
Roughly speaking, an \emph{admissible transformation} in a class of DEs
is a~transformation that maps an equation from this class
to another equation from the same class.
Therefore, an admissible transformation can be regarded as a triple consisting of
an initial equation,
a~target equation
and a~point transformation that links them.
A~number of group classification problems were recently solved using such transformations.
See, e.g.,~\cite{bihl2012b,popo2010a,popo2010c,vane2009a,vane2012b} and references therein
for related notions, techniques and results.

The structure of the paper is as follows.
In Section~\ref{Section_AT} we start the study of Lie symmetries of the class~\eqref{GBE}
with the description of its equivalence groupoid
in terms of normalized subclasses and conditional equivalence groups of different kinds.
Using parameterized families of transformations from the equivalence group of the entire class~\eqref{GBE},
we can gauge arbitrary elements of the class, which generates mappings of this class onto its subclasses.
This is done for the gauge $h=0$ in Section~\ref{Section_gauge}.
Lie symmetries of the class~\eqref{GBE} are classified in Section~\ref{Section_LS}
via reducing the problem to the group classification of the subclass singled out by the constraint $h=0$.
In Section~\ref{SectionISA} we find optimal systems of one- and two-dimensional subalgebras
for all inequivalent cases of Lie symmetry extensions in the class~\eqref{GBE}.
Lie reductions with respect to subalgebras from the optimal systems are performed in Section~\ref{SectionReduction},
as well as some exact solutions to equations from the class~\eqref{GBE} with $h=0$ are constructed.
Section~\ref{SectionGeneration} is devoted to the generation of exact solutions to such equations with $h\ne0$.
The linearization of equations from the class~\eqref{GBE} to the heat equation is discussed in Section~\ref{Section_LGBE}.

\section{Admissible transformations}
\label{Section_AT}

In order to study the admissible transformations in the class~\eqref{GBE}, we suppose
that an equation of the form~\eqref{GBE} is connected with an equation
\begin{gather}
\label{GBE_tilde}
{\tilde u}_{\tilde t}+\tilde u^{\tilde n}{\tilde u}_{\tilde x}+\tilde h(\tilde t)\tilde u=
\tilde g(\tilde t){\tilde u}_{\tilde x\tilde x}
\end{gather}
from the same class by a~point transformation
$\tilde t=T(t,x,u)$,
$\tilde x=X(t,x,u)$,
$\tilde u=U(t,x,u)$,
where $|\partial(T,X,U)/\partial(t,x,u)|\ne0$.
It is known that admissible transformations of evolution equations obey the restrictions $T_x=T_u=0$~\cite{king1998a}.
Moreover, each admissible transformation between equations of the form~$u_t=F(t,x,u)u_{xx}+G(t,x,u,u_x)$
necessarily satisfies the condition $X_u=0$~\cite{popo2004b}.
Therefore, for the class~\eqref{GBE} it suffices to consider transformations of the form
\begin{gather*}
\tilde t=T(t),
\quad
\tilde x=X(t,x),
\quad
\tilde u=U(t,x,u),
\end{gather*}
where $T_tX_xU_u\ne0$.
Expressing all tilded entities in~\eqref{GBE_tilde} in terms of the untilded variables,
we substitute $u_t=-u^nu_x-h(t)u+g(t)u_{xx}$ into the rewritten equation~\eqref{GBE_tilde}
in order to confine it to the manifold defined by~\eqref{GBE} in the second-order jet space
with the independent variables $(t,x)$ and the dependent variable~$u$.
The splitting of the obtained identity with respect to the derivatives $u_{xx}$ and $u_x$ leads to the determining
equations on the functions~$T$, $X$ and $U$,
\begin{gather}\arraycolsep=0ex
\begin{array}{l}\label{dequations}
U_{uu}=0,
\quad
\tilde g T_t=g X_x^2,
\\[1ex]
u^{n}-\dfrac{T_t}{X_x}U^{\tilde n}
+2\tilde g\dfrac{T_t}{X_x^2}\dfrac{U_{xu}}{U_u}-\tilde g T_t\dfrac{X_{xx}}{X_x^3}+\dfrac{X_t}{X_x}=0,
\\[2ex]
X_t U_x-X_x U_t+\tilde g T_t\left(\dfrac{U_x}{X_x}\right)_x\!-T_t U^{\tilde n}U_x+h uX_xU_u-\tilde h T_t X_x U=0.
\end{array}
\end{gather}
The first equation implies that $U$ is linear in $u$, $U=U^1(t,x)u+U^0(t,x)$.
From the second one we derive that $X_x$ is a~function only of $t$, so $X=X^1(t)x+X^0(t)$, and
\begin{gather*}
\tilde g=\frac{(X^1)^2}{T_t}\,g.
\end{gather*}
Differentiating the third equation with respect to $u$,
we prove that the arbitrary element~$n$ is invariant under the action of admissible point
transformations, i.e., $\tilde n=n$, and, moreover, if $n\ne1$, then $U^0=0$.
The further consideration depends on whether $n\ne1$ or $n=1$.
Substituting the expressions for $U$, $X$ and $\tilde g$ into the third and the fourth determining equations,
we split them with respect to $u$ and $x$.

\paragraph{Case~$\boldsymbol{n\ne1}$.} The splitting implies $U^1_x=0$, $X^1_t=0$, $X^0_t=0$.
Solving the remaining determining equations,
$\left(U^1\right)^nT_t=X^1$ and $\tilde h X^1\big(U^1\big)^{1-n}=hU^1-U^1_t$,
we get the transformation components for~$u$ and~$h$.
The transformations obtained are parameterized by the constant arbitrary element~$n$,
and they can be applied to all equations from the class~\eqref{GBE} including those with $n=1$.
Therefore, these transformations form the generalized equivalence group $\hat G^\sim$ of the class~\eqref{GBE},
which leads to the following statement.

\begin{theorem}\label{Theorem_GenBurgers_G}
The generalized equivalence group $\hat G^\sim$ of the class~\eqref{GBE} consists of the transformations
\begin{gather*}
\tilde t=T(t),
\quad
\tilde x=\delta_1x+\delta_0,
\quad
\tilde u=\left(\dfrac{\delta_1}{T_t}\right)^\frac1nu,
\quad
\tilde h=\frac1{T_t}{h}+\frac{T_{tt}}{nT_t^{\,2}},
\quad
\tilde g=\frac{{\delta_1}^2}{T_t}\,g,
\quad
\tilde n=n,
\end{gather*}
where $\delta_1$ and $\delta_0$ are arbitrary constants and $T=T(t)$ is an arbitrary smooth function with $\delta_1T_t>0$.
The equivalence groupoid of the subclass of the class~\eqref{GBE} singled out
by the condition $n\ne1$ is generated by elements of~$\hat G^\sim$,
i.e., this subclass is normalized in the generalized sense.
\end{theorem}

\begin{remark}
If we assume that the power~$n$ varies in the class~\eqref{GBE}, then the equivalence group in
Theorem~\ref{Theorem_GenBurgers_G} is generalized since $n$ is involved in the transformation of the dependent variable~$u$.
The notion of usual equivalence group supposes
that transformation components for the independent and dependent variables do not depend on the arbitrary elements.
From the other hand, $n$ is invariant under the action of transformations from the equivalence group,
so the class~\eqref{GBE} can be considered as the union of its subclasses with fixed~$n$.
Under fixing~$n$ the group $\hat G^\sim$ generates the usual equivalence groups for these subclasses.
\end{remark}

\begin{remark}\looseness=-1
The signs of bases in powers with exponents containing $n$ should be carefully treated throughout the paper.
Thus, solutions of an equation from the class~\eqref{GBE} should be positive whenever the exponent~$n$ is not a rational number with odd denominator.
Therefore, the condition $u>0$ is natural when the entire class~\eqref{GBE} with varying~$n$ is considered.
Moreover, usually the positivity of~$u$ well agrees with the physical meaning of this function.
In order to avoid the positivity restriction on~$u$ we could formally replace~$u^n$ in~\eqref{GBE} by $|u|^n$.
If we study a subclass of~\eqref{GBE} with a fixed exponent~$n$ being a rational number with odd numerator and odd denominator,
then we can omit the positivity restriction on~$u$
and equivalence transformations with $\delta_1T_t<0$ become allowed,
i.e., we can alternate the signs of pairs~$(t,u)$ and~$(x,u)$ independently.
We neglect such transformations in the paper.
\end{remark}

Theorem~\ref{Theorem_GenBurgers_G} allows us to easily derive
the conditional equivalence group of~\eqref{GBE} for the case $h=\const$,
which was considered in~\cite{vaga2006a}.
The transformation component for the arbitrary element $h$,
\begin{gather*}
\tilde h=\frac1{T_t}{h}+\frac{T_{tt}}{nT_t^{\,2}},
\end{gather*}
can be treated as an equation in the parameter-function $T$.
We present the general solution of this equation in a~form that brings out
the continuous dependence on the parameters~$h$ and~$\tilde h$.

\begin{corollary}
The generalized equivalence group~$\hat G^\sim_{h=\const}$ of the class~\eqref{GBE} with $h=\const$ consists of the
transformations
\begin{gather*}
\textstyle\tilde t=T(t),
\quad
\tilde x=\delta_1x+\delta_0,
\quad
\tilde u=\left(\dfrac{\delta_1}{\alpha}\right)^{\frac1n}e^{ht-\tilde h T}u,
\quad
\tilde g=\dfrac{{\delta_1}^2}{\alpha}e^{nht-n\tilde hT}g,
\quad
\tilde n=n,
\end{gather*}
where the function $T=T(t)$ depends on $h$ and $\tilde h$ and is defined by the formulae
\begin{gather*}
\arraycolsep=0ex
\begin{array}{ll}
h\tilde h\not=0\colon
\quad
\dfrac{e^{-n\tilde h T}-1}{-n\tilde h}=\alpha\dfrac{e^{-n h t}-1}{-n h}+\beta,
\qquad
&h\not=0,\;\tilde h=0\colon
\quad
T=\alpha\dfrac{e^{-n h t}-1}{-n h}+\beta,
\\[3.5mm]
h=0,\;\tilde h\not=0\colon
\quad
\dfrac{e^{-n\tilde h T}-1}{-n\tilde h}=\alpha t+\beta,
\qquad
&h=\tilde h=0\colon
\quad
T=\alpha t+\beta.
\end{array}
\end{gather*}
Here $\alpha$, $\beta$, $\delta_0$ and $\delta_1$ are arbitrary constants with $\alpha\delta_1>0$.
\end{corollary}

\vspace*{-4mm}

\paragraph{Case~$\boldsymbol{n=1}$.} The splitting of the third and the fourth equations~\eqref{dequations}
leads to the system
\begin{gather}
U^1_x=0,
\quad
U^1=\frac{X^1}{T_t},
\quad
U^0=\frac{X^1_tx+X^0_t}{T_t},
\quad
\tilde h=\frac1{T_t}\left(h+\frac{T_{tt}}{T_t}-2\frac{X^1_t}{X^1}\right), \label{a_tildea}
\\
X^1_{tt}-\frac{2\big(X^1_t\big)^2}{X^1}+h X^1_t=0,
\quad
X^0_{tt}-\frac{2X^0_tX^1_t}{X^1}+h X^0_t=0.
\nonumber
\end{gather}
The second-order equations on $X^1$ and $X^0$ imply $X^0=\delta_1X^1+\delta_0$, where
\begin{gather}
\label{X1}
X^1=\left(\gamma\int e^{-\int h(t)\,\D t}\D t+\delta\right)^{-1}\!,
\end{gather}
and $\delta_0$, $\delta_1$, $\delta$ and $\gamma$ are arbitrary constants with $(\gamma,\delta)\ne(0,0)$.
Here and in what follows an integral with respect to~$t$ should be interpreted as a fixed antiderivative.

\begin{theorem}
\label{Theorem_GenBurgers_n1_G}
The generalized extended equivalence group $\hat G^\sim_{n=1}$ of the class
\begin{gather}
\label{GBE_n1}
u_t+uu_x+h(t)u=g(t)u_{xx}
\end{gather}
consists of the transformations
\begin{gather*}
\tilde t=T(t),
\quad
\tilde x=(x+\delta_1)X^1+\delta_0,
\quad
\tilde u=\frac{X^1}{T_t}\left(u+(x+\delta_1)\frac{X^1_t}{X^1}\right),
\\
\tilde h=\frac1{T_t}\left(h+\frac{T_{tt}}{T_t}-2\frac{X^1_t}{X^1}\right),
\quad
\tilde g=\frac{\big(X^1\big)^2}{T_t}g.
\end{gather*}
Here $\delta_0$, $\delta_1$ are arbitrary constants, $T=T(t)$ is an arbitrary smooth function with $T_t\ne0$, and the
function $X^1=X^1(t)$ is defined by~\eqref{X1}.
Moreover, this class is normalized in the generalized extended sense.
\end{theorem}

In order to complete the study of equivalence transformations between equations
with constant value of the arbitrary element~$h$,
we consider the subclass of the class~\eqref{GBE_n1} singled out by the constraint $h=\const$.
In this case the parameter $T=T(t)$ is not an arbitrary function but a~solution
of the last equation in~\eqref{a_tildea} with $X^1$ defined by the formulae
\begin{gather}\label{x1}
X^1=
\begin{cases}
\left(\gamma\dfrac{e^{-h t}-1}{-h}+\delta\right)^{-1}&\text{if}
\quad
h\ne0,
\\[2.8mm]
\left(\gamma t+\delta\right)^{-1}&\text{if}
\quad
h=0,
\end{cases}
\end{gather}
where $\gamma$, $\delta$ are arbitrary constants, $(\gamma,\delta)\ne(0,0)$.

Integrating the last equation in~\eqref{a_tildea} we get the following statement.
\begin{corollary}
The generalized equivalence group~$\hat G^\sim_{n=1,\,h=\const}$ of the class~\eqref{GBE_n1} with $h=\const$
consists of the transformations
\begin{gather*}
\tilde t=T(t),
\quad
\tilde x=(x+\delta_1)X^1+\delta_0,
\quad
\tilde u=\frac{e^{h t-\tilde h T}\!}{\Delta X^1}\!\left(u+(x+\delta_1)\frac{X^1_t}{X^1}\right)\!,
\quad
\tilde g=\frac1{\Delta}{e^{h t-\tilde h T}}g,
\end{gather*}
where the function $X^1$ is of form~\eqref{x1}
and the function $T=T(t)$ is defined by the formulae
\begin{gather*}
\arraycolsep=0ex
\begin{array}
{ll}h\tilde h\not=0\colon
\quad
\dfrac{e^{-\tilde h T}-1}{-\tilde h}=\dfrac{\alpha\dfrac{e^{-h t}-1}{-h}+\beta}{\gamma\dfrac{e^{-h t}-1}{-h}+\delta},
\qquad
&h\not=0,\;\tilde h=0\colon
\quad
T=\dfrac{\alpha\dfrac{e^{-h t}-1}{-h}+\beta}{\gamma\dfrac{e^{-h t}-1}{-h}+\delta},
\\[7mm]
h=0,\;\tilde h\not=0\colon
\quad
\dfrac{e^{-\tilde h T}-1}{-\tilde h}=\dfrac{\alpha t+\beta}{\gamma t+\delta},
\qquad
&h=\tilde h=0\colon
\quad
T=\dfrac{\alpha t+\beta}{\gamma t+\delta}.
\end{array}
\end{gather*}
Here $\alpha$, $\beta$, $\gamma$ and $\delta$ are arbitrary constants defined up to a~nonzero multiplier,
$\Delta=\alpha\delta-\beta\gamma\ne0$.
\end{corollary}

Therefore, the admissible transformations in the class~\eqref{GBE} are exhaustively described.
The following statement is true.

\begin{theorem}\label{Theorem_GenBurgers_EquivGroupoid}
The class~\eqref{GBE}, where the exponent~$n$ varies, is not normalized.
It can be partitioned into the normalized subclasses each of which is singled out by fixing a value of $n$,
and these classes are not connected by point transformations.
Each subclass of~\eqref{GBE} with a~specified value of~$n$, $n\ne1$, is normalized in the usual sense,
whereas the subclass~\eqref{GBE_n1} corresponding to the value $n=1$
is normalized in the generalized extended sense only.
Every union of any subclasses of~\eqref{GBE} with $n\ne1$ is normalized in the generalized sense.
\end{theorem}

\section{Gauging the arbitrary elements}
\label{Section_gauge}

The transformations from $\hat G^\sim$ are parameterized by an arbitrary function $T=T(t)$.
This allows us to gauge one of the arbitrary elements, $g$ or $h$, to a~simple constant value.
For example, we can set $g$ to one or $h$ to zero.
The gauge $h=0$ looks more convenient since in this case the class~\eqref{GBE} reduces to another one,
for which the group classification problem has been solved recently~\cite{vane2013b}.%
\footnote{
Group classification problems for certain subclasses
of~\eqref{GBE} with $h=0$ were considered in~\cite{doyl1990a,wafo2004d,vaga2009b,vaga2006a}.
}
This gauge can be realized with the transformation
\begin{gather}\label{tr_tau}
\mathcal T\colon
\quad
\hat t=\int e^{-n\int h(t)\D t}\D t,
\quad
\hat x=x,
\quad
\hat u=e^{\int h(t)\D t}u
\end{gather}
from the group~$\hat G^\sim$, which links the class~\eqref{GBE} with the class
$\hat u_{\hat t}+\hat u^n\hat u_{\hat x}=\hat g(\hat t)\hat u_{\hat x\hat x}$,
where the new arbitrary element $\hat g$ depends on $h$ and $g$ as
\begin{gather}
\label{tr_g}
\hat g=e^{\,n\!\int h(t)\,\D t}g.
\end{gather}

Of course the transformation $\mathcal T$ for this gauging is not unique.
If $n\ne1$, then the most general transformation is
\begin{gather*}
\hat t=\alpha\int e^{-n\int h(t)\D t}\D t+\beta, \quad
\hat x=\delta_1x+\delta_0, \quad
\hat u=\left(\frac{\delta_1}{\alpha}\right)^{\frac1n}e^{\int h(t)\D t}u,
\end{gather*}
where $\alpha$, $\beta$, $\delta_1$ and $\delta_0$ are arbitrary constants with $\alpha\delta_1\neq0$.
In the case $n=1$ the most general transformation takes the form
\begin{gather*}
\tilde t=\dfrac{\alpha\hat t+\beta}{\gamma\hat t+\delta}, \quad
\tilde x=\dfrac{x+\mu_1\hat t+\mu_0}{\gamma\hat t+\delta}, \quad
\tilde u=\frac{(\gamma\hat t+\delta)e^{\int h(t)\D t}u-\gamma x+\mu_1\delta-\mu_0\gamma}{\alpha\delta-\beta\gamma}, \quad
\end{gather*}
where $\alpha$, $\beta$, $\gamma$, $\delta$, $\mu_0$, and $\mu_1$
are arbitrary constants with $\alpha\delta-\beta\gamma\ne0$,
and $\hat t=\int e^{-n\int h(t)\D t}\D t$.

If $h$ is a nonzero constant, then the transformation $\mathcal T$ gauging $h$ to zero has the form
\begin{gather*}
\hat t=-\frac1{nh}e^{-nht}, \quad
\hat x=x, \quad
\hat u=e^{h t}u.
\end{gather*}

Theorems~\ref{Theorem_GenBurgers_G}--\ref{Theorem_GenBurgers_EquivGroupoid}
exhaustively describe the equivalence groupoid of the class~\eqref{GBE}.
This allows us to easily find the equivalence groupoid of any subclass of the class~\eqref{GBE}.
Consider the two gauged classes that respectively consist of the equations
\begin{gather}
u_t+u^nu_x=g(t)u_{xx}, \quad ng\ne0,  \label{GBE_h0}\\
u_t+uu_x=g(t)u_{xx},\quad g\ne0.      \label{GBE_n1_h0}
\end{gather}

\begin{corollary}\label{GenBurgers_G_h0_Cor}
The subclass of the class~\eqref{GBE_h0} singled out by the constraint $n\ne1$
is normalized in the generalized sense.
More precisely, the equivalence groupoid of this class is generated by
its generalized equivalence group,
which coincides with the generalized equivalence group~$\hat G^\sim_{h=0}$
of the entire class~\eqref{GBE_h0} and consists of the transformations
\begin{gather*}
\textstyle\tilde t=\alpha t+\beta, \quad
\tilde x=\delta_1x+\delta_0, \quad
\tilde u=\left(\dfrac{\delta_1}{\alpha}\right)^\frac1n u, \quad
\tilde g(\tilde t)=\dfrac{\delta_1^2}{\alpha}g(t), \quad
\tilde n=n,
\end{gather*}
where $\alpha$, $\beta$, $\delta_0$ and $\delta_1$ are arbitrary constants with $\alpha\delta_1>0$.

\end{corollary}

\begin{corollary}\label{GenBurgers_n1_G_h0_Cor}
The class~\eqref{GBE_n1_h0}, which is singled out from the class~\eqref{GBE_h0} by the constraint \mbox{$n=1$},
is normalized in the usual sense, and
its usual equivalence group $\hat G^\sim_{h={\rm0},\,n=1}$ consists of the transformations
\begin{gather*}
\tilde t=\dfrac{\alpha t+\beta}{\gamma t+\delta},
\quad
\tilde x=\dfrac{x+\mu_1t+\mu_0}{\gamma t+\delta},
\quad
\tilde u=\frac{(\gamma t+\delta)u-\gamma x+\mu_1\delta-\mu_0\gamma}{\alpha\delta-\beta\gamma},
\quad
\tilde g=\frac1{\alpha\delta-\beta\gamma}g.
\end{gather*}
Here $\alpha$, $\beta$, $\gamma$, $\delta$, $\mu_0$, and $\mu_1$ are arbitrary constants with
$\alpha\delta-\beta\gamma\ne0$.\footnote{
The group~$\hat G^\sim_{h={\rm0},\,n=1}$ was found earlier in~\cite{king1991c,poch2012a,poch2013b}
in the course of study of admissible transformations in the class of generalized Burgers equations
$u_t+uu_x+g(t,x)u_{xx}=0$ and its subclass with $g=g(t)$.
}
\end{corollary}

\begin{remark}
The alternative gauge $g=1$ can be set using a parameterized family of point transformations that are projections of transformations from~$\hat G^\sim$
to the space of independent and dependent variables,
\begin{gather*}
\hat t=\int g(t)\D t,
\quad
\hat x=x\sgn g(t),
\quad
\hat u=|g(t)|^{-\frac1n}u.
\end{gather*}
This family of transformations maps the class~\eqref{GBE} onto the class
$\hat u_{\hat t}+\hat u^n\hat u_{\hat x}+\hat h(\hat t)\hat u=\hat u_{\hat x\hat x}$,
where the new arbitrary element $\hat h$ depends on $h$ and $g$ as
$\hat h=\frac hg+\frac{g_t}{ng^2}$.
\end{remark}

\section{Lie symmetries} \label{Section_LS}

In the previous section we have proven that the group classification problem for the class~\eqref{GBE}
can be reduced to the similar problem for the subclass singled out by the condition $h(t)=0$,
i.e., for the class~\eqref{GBE_h0}.
The classical approach~\cite{Olver1986,ovsy1982a} to finding Lie symmetries
prescribes to look for vector fields of the form $Q=\tau(t,x,u)\partial_t+\xi(t,x,u)\partial_x+\eta(t,x,u)\partial_u$
that generate one-parameter groups of point symmetry transformations of an equation from the class~\eqref{GBE_h0}.
Any such vector field~$Q$ meets the infinitesimal invariance criterion, i.e.,
the action of the second prolongation~$Q^{(2)}$ of~$Q$ on the left hand side of~\eqref{GBE_h0}
results in the expression identically satisfied by all solutions of this equation, which reads
\begin{equation}\label{EqInfinitesimalInvCriterionForGBE}
Q^{(2)}\{u_t+u^nu_x-g(t)u_{xx}\}\Big|_{u_t=-u^nu_x+g(t)u_{xx}}=0.
\end{equation}
Here $Q^{(2)}=Q+\eta^t\partial_{u_t}+\eta^x\partial_{u_x}+\eta^{xx}\partial_{u_{xx}}$, where
\begin{gather*}
\eta^t=D_t(\eta)-u_tD_t(\tau)-u_xD_t(\xi),\\
\eta^x=D_x(\eta)-u_tD_x(\tau)-u_xD_x(\xi),\\
\eta^{xx}=D_x(\eta^x)-u_{tx}D_x(\tau)-u_{xx}D_x(\xi),
\end{gather*}
$D_t=\partial_t+u_t\partial_{u}+u_{tt}\partial_{u_t}+u_{tx}\partial_{u_x}+\dots{}$ and
$D_x=\partial_x+u_x\partial_{u}+u_{tx}\partial_{u_t}+u_{xx}\partial_{u_x}+\dots{}$
are the total derivatives with respect to~$t$ and~$x$, respectively.
The condition~\eqref{EqInfinitesimalInvCriterionForGBE}
leads to the determining equations in the coefficients~$\tau$, $\xi$ and $\eta$.
The simplest determining equations immediately imply
\begin{gather*}
\tau=\tau(t),\quad
\xi=\xi(t,x), \quad
\eta=\eta^1(t,x)u+\eta^0(t,x),
\end{gather*}
where $\tau$, $\xi$, $\eta^1$ and $\eta^0$ are arbitrary smooth functions of their variables.
The remaining determining equations are
\begin{gather*}
2g\xi_x=(g\tau)_t,
\\
\eta^1_xu^{n+1}+\eta^0_xu^{n}+(\eta^1_t-g\eta^1_{xx})u+\eta^0_t-g\eta^0_{xx}=0,
\\
(\tau_t-\xi_x+n\eta^1)u^n+n\eta^0u^{n-1}+g\xi_{xx}-2g\eta^1_x-\xi_t=0.
\end{gather*}
The splitting of the second and the third equations
with respect to $u$ depends on the arbitrary element~$n$.
Only the case $n=1$ appears to differ from the other cases of the splitting.
The determining equations in this case become
\begin{gather*}
2g\xi_x=(g\tau)_t,\quad \eta^1_x=0,\quad \eta^0_x+\eta^1_t=0,\quad \tau_t-\xi_x+\eta^1=0,\quad \eta^0-\xi_t=0.
\end{gather*}
For the other nonzero values of~$n$ the coefficient~$\eta^0$ equals zero, and the determining equations take the form
\begin{gather*}
2g\xi_x=(g\tau)_t,\quad \eta^1_x=0,\quad \tau_t-\xi_x+n\eta^1=0,\quad \xi_t=0.
\end{gather*}

\begin{table}[ht!]
\small
\begin{center}
\refstepcounter{table}\renewcommand{\arraystretch}{1.65}
\textbf{Table~\thetable.} The group classification of the class~\eqref{GBE} up to the general point equivalence.
\\[1.2ex]
\begin{tabular}
{|c|c|l|}
\hline
no.
&
$g$
&
\hfil Basis operators of $A^{\rm max}$
\\
\hline
\multicolumn{3}{|c|}{${n\ne1}$. This case is classified up to $\hat G^\sim$-equivalence.}
\\
\hline
1
&
$\forall$
&
$\partial_x$
\\
\hline
2
&
$\varepsilon t^\rho$
&
$\partial_x,
\quad
2nt\partial_t+n(\rho\!+\!1\!)x\partial_x+(\rho\!-\!1\!)u\partial_u$
\\
\hline
3
&
$\varepsilon e^t$
&
$\partial_x,
\quad
2n\partial_t+nx\partial_x+u\partial_u$
\\
\hline
4
&
$1$
&
$\partial_x,
\quad
\partial_t,
\quad
2nt\partial_t+nx\partial_x-u\partial_u$
\\
\hline
\multicolumn{3}{|c|}{$n=1$. This case is classified up to $\hat G^\sim_{n=1}$-equivalence.}
\\
\hline
5
&
$\forall$
&
$\partial_x,
\quad
t\partial_x+\partial_u$
\\
\hline
6
&
$\varepsilon t^\rho$
&
$\partial_x,
\quad
t\partial_x+\partial_u,
\quad
2t\partial_t+(\rho\!+\!1\!)x\partial_x+(\rho\!-\!1\!)u\partial_u$
\\
\hline
7
&
$\varepsilon e^t$
&
$\partial_x,
\quad
t\partial_x+\partial_u,
\quad
2\partial_t+x\partial_x+u\partial_u$
\\
\hline
8
&
$\varepsilon e^{2\rho\arctan t}$
&
$\partial_x,
\quad
t\partial_x+\partial_u,
\quad
(t^2+1)\partial_t+(t+\rho)x\partial_x+(x+(\rho-t)u)\partial_u$
\\
\hline
9
&
$1$
&
$\partial_x,
\quad
t\partial_x+\partial_u,
\quad
\partial_t,
\quad
2t\partial_t+x\partial_x-u\partial_u,
\quad
t^2\partial_t+tx\partial_x+(x-tu)\partial_u$
\\
\hline
\end{tabular}
\end{center}
Here $h(t)=0\bmod\hat G^\sim$, $\varepsilon=\pm1\bmod\hat G^\sim$,
$\rho$ is a~nonzero constant.
In Cases~6 and~8 we can set, $\bmod~\hat G^\sim_{n=1}$, either~$\rho>0$ or $\rho<0$.
\par\medskip
\begin{center}
\refstepcounter{table}\renewcommand{\arraystretch}{1.65}
\textbf{Table~\thetable.} The complete list of Lie symmetry extensions for the class~\eqref{GBE_h0}.
\\[1.2ex]
\begin{tabular}
{|c|c|l|}
\hline
no.
&
${g}$
&
\hfil Basis operators of $A^{\rm max}$
\\
\hline
\multicolumn{3}{|c|}{$n\ne1$}
\\
\hline
1
&
$\forall$
&
$\partial_x$
\\
\hline
2
&
$\lambda(\alpha t+\beta)^\rho$
&
$\partial_x,
\quad
2n(\alpha t+\beta)\partial_t+\alpha n(\rho\!+\!1\!)x\partial_x+\alpha(\rho\!-\!1\!){u}\partial_{u}$
\\
\hline
3
&
$\lambda e^{\alpha t}$
&
$\partial_x,
\quad
2n\partial_t+\alpha nx\partial_x+\alpha{u}\partial_{u}$
\\
\hline
4
&
$\lambda$
&
$\partial_t,
\quad
\partial_x,
\quad
2nt\partial_t+nx\partial_x-{u}\partial_{u}$
\\
\hline
\multicolumn{3}{|c|}{$n=1$}
\\
\hline
5
&
$\forall$
&
$\partial_x,
\quad
t\partial_x+\partial_{u}$
\\
\hline
6
&
$\lambda\left(\frac{\alpha t+\beta}{\gamma t+\delta}\right)^\rho$
&
$\partial_x,
\quad
t\partial_x+\partial_{u}$, \\[-2mm]&&
$(\alpha t+\beta)(\gamma t+\delta)\partial_t+\left(\frac12(\rho\!-\!1\!)\Delta+\alpha(\gamma t+\delta)\right)x\partial_x$
\\[-2mm]&&
$\quad+\left(\left[\frac12(\rho\!+\!1\!)\Delta-\alpha(\gamma t+\delta)\right]{u}+\alpha\gamma x\right)\partial_{u}$
\\
\hline
7
&
$\lambda e^{\frac{\alpha t+\beta}{\gamma t+\delta}}$
&
$\partial_x,
\quad
t\partial_x+\partial_{u}$, \\[-2mm]&&
$(\gamma t+\delta)^2\partial_t+\left(\frac12\Delta+\gamma(\gamma t+\delta)\right)
x\partial_x+\left(\left[\frac12\Delta-\gamma(\gamma t+\delta)\right]{u}
+\gamma^2x\right)\partial_{u}$
\\
\hline
8
&
$\lambda e^{2\rho\arctan(\alpha t+\beta)}$
&
$\partial_x,
\quad
t\partial_x+\partial_{u}$, \\[-2mm]&&
$\left((\alpha t+\beta)^2+1\right)\partial_t+\alpha\left(\alpha t+\rho+\beta\right)x\partial_x+\alpha\left(\left[-\alpha t+ \rho-\beta\right]
{u}+\alpha x\right)\partial_{u}$
\\
\hline
9
&
$\lambda$
&
$\partial_t,
\quad
\partial_x,
\quad
t\partial_x+\partial_{u},
\quad
2t\partial_t+x\partial_x-{u}\partial_{u},
\quad
t^2\partial_t+tx\partial_x+(x-t{u})\partial_{u}$
\\
\hline
\end{tabular}
\end{center}
Here $\lambda$ and $\rho$ are nonzero constants.
$\alpha=\pm1$ in Case~2 and $\alpha\ne0$ in Cases~3 and 8.
In Cases~6 and~7 $\alpha$,~$\beta$, $\gamma$, and~$\delta$ are arbitrary constants defined up to a nonzero multiplier
(with additional possibility of scaling in Case~6)
such that $\Delta=\alpha\delta-\beta\gamma\ne0$.
In Case~7 we can set $(\alpha,\beta,\gamma,\delta)\in\{(\alpha',0,0,1),(0,\beta',1,\delta')\}$.

\vspace{-4ex}
\end{table}

The group classification of the class~\eqref{GBE} up to the general point equivalence,
which is in fact generated in this class by the equivalence groups
from Theorems~\ref{Theorem_GenBurgers_G} and~\ref{Theorem_GenBurgers_n1_G},
is presented in Table~1.
This classification can also be interpreted as the group classification of the class~\eqref{GBE_h0}
up to the general point equivalence, which is generated in the class~\eqref{GBE_h0} by
the equivalence groups given in Corollaries~\ref{GenBurgers_G_h0_Cor} and~\ref{GenBurgers_n1_G_h0_Cor}.%
\footnote{The group classification problem for the class of equations of the form
$u_t+a(u^m)_x=g(t)u_{xx}$ with $g\ne0$, $ m\ne0,1$ and $a=\const\ne0$,
which is in fact another representation of the subclass of~\eqref{GBE_h0} with $n\neq-1$ under setting $a=1/m$ and $m=n+1$,
was solved in~\cite{vane2013b}. Hence the classification lists for these classes can be derived from each other.
As the case $n=-1$ is not singular from the Lie symmetry point of view,
we do not exclude this value of~$n$ from the classification list for the class~\eqref{GBE_h0}.}
The intersection of the maximal Lie invariance algebras of equations
from the class~\eqref{GBE} (resp.\ from the class~\eqref{GBE_h0}) is the algebra $\langle\partial_x\rangle$,
which is given as Case~1 of Tables~1--3.
\looseness=-1

Table~2 contains the complete list of Lie symmetry extensions for the class~\eqref{GBE_h0}, where
the forms of the arbitrary element $g$ are not simplified by equivalence transformations.%
\footnote{
The group classification of the class~\eqref{GBE_n1_h0},
which is a subclass of the class~\eqref{GBE_h0} singled out by the constraint $n=1$,
was carried out in~\cite{doyl1990a,wafo2004d} with weaknesses.
Thus, in~\cite{doyl1990a} each of Cases 6 and 7 of Table~2
was multiplied twofold,
$g=e^{\alpha t}$, $g=e^{\frac1{\alpha t}}$
and $g=(\alpha t+\beta)^{\rho}$, $g=\left(\frac{\alpha t+\beta}{\gamma t+\delta}\right)^\rho$,
respectively.
The similar weaknesses are contained in~\cite{wafo2004d}.
The corresponding equivalence group was neither computed nor utilized in these papers.}
There are some remarks on Cases~7 and~8 of this table.

Only three constants are really independent in Case 7.
To show this, we consider two subcases depending on whether $\gamma$ is zero or not.
If $\gamma=0$, then $ g =\lambda'\exp(\alpha' t)$, where $\lambda'=\lambda\exp(\beta/\delta)$, $\alpha'=\alpha/\delta$.
If $\gamma\neq0$, then $ g =\lambda'\exp\big(\frac{\beta'}{t+\delta'}\big)$,
where $\lambda'=\lambda\exp(\alpha/\gamma)$, $\beta'=(\beta\gamma-\alpha\delta)/\gamma^2$, $\delta'=\delta/\gamma$.

Case~8 seems not to be maximally general but this impression is deceptive.
Indeed, similarly to Cases~6 and~7 we can consider the value
$g=\lambda\exp\!\Big({2\rho\arctan\!\frac{\alpha t+\beta}{\gamma t+\delta}}\Big)$
with $\Delta=\alpha\delta-\beta\gamma\neq0$.
The corresponding Lie symmetry algebra is spanned by the basis operators
\begin{gather}\nonumber
\partial_x,
\quad
t\partial_x+\partial_u,
\quad
\left((\alpha t+\beta)^2+(\gamma t+\delta)^2\right)\partial_ t
+\left((\alpha^2+\gamma^2) t+\rho\Delta+\alpha\beta+\gamma\delta\right)x\partial_x
\\\label{EqSymOpsOfGBEwithArctan}
\qquad
+\left(\left[-(\alpha^2+\gamma^2) t+\rho\Delta-\alpha\beta-\gamma\delta\right]
{u}+(\alpha^2+\gamma^2)x\right)\partial_{u}.
\end{gather}
However, the formula for sum of arctangents,
\begin{gather*}
\arctan y+\arctan z=\left\{\begin{array}{ll}
\arctan\dfrac{y+z}{1-yz}+\dfrac\pi2(\sgn y)(1+\sgn(1-yz)),\quad&  \mbox{if}\quad yz\ne1,\\
\dfrac\pi2\sgn y,&  \mbox{if}\quad yz=1,
\end{array}
\right.
\end{gather*}
implies that the above value of $g$ locally coincides with
$\check g=\check\lambda\exp(2\check\rho\arctan(\check\alpha t+\check\beta))$.
Here
\begin{gather*}
\check\lambda=\lambda e^{-2\rho\arctan\frac\gamma\alpha
-\pi\rho\sgn\big(1+\frac\gamma\alpha(\check\alpha t+\check\beta)\big)\sgn\frac\gamma\alpha},
\quad
\check\rho=\rho,
\quad
\check\alpha=\frac{\alpha^2+\gamma^2}{\alpha\delta-\beta\gamma},
\quad
\check\beta=\frac{\alpha\beta+\gamma\delta}{\alpha\delta-\beta\gamma}
\end{gather*}
for $\gamma\neq0$, $\alpha\neq0$, and
\begin{gather*}
\check\lambda=\lambda e^{\pi\rho\sgn(\check\alpha t+\check\beta)},
\quad
\check\rho=-\rho,
\quad
\check\alpha=-\frac{\gamma}{\beta},
\quad
\check\beta=-\frac{\delta}{\beta}
\end{gather*}
for $\gamma\neq0$, $\alpha=0$.
The case $\alpha\neq0$, $\gamma=0$ is obvious.
The above expressions for constant parameters also agree with the simplified
form of the third operator in~\eqref{EqSymOpsOfGBEwithArctan}.

To derive the complete list of Lie symmetry extensions for the entire class~\eqref{GBE},
where arbitrary elements are not simplified by point transformations, we use the equivalence-based approach.%
\footnote{This approach was successfully applied to deriving
the complete classification lists for certain classes of variable coefficient KdV and mKdV equations in~\cite{vane2012a,vane2013a}.}
For this purpose we apply the transformation $\mathcal T$ given by~\eqref{tr_tau} to the vector fields presented in Table~2,
and find the corresponding values of the arbitrary element~$g$ by means of~\eqref{tr_g}. The results are collected in Table~3.

\begin{table}[h!]\small
\begin{center}
\refstepcounter{table}\renewcommand{\arraystretch}{1.6}  
\textbf{Table~\thetable.} The complete list of Lie symmetry extensions for the class~\eqref{GBE}.
\\[1.2ex]
\begin{tabular}
{|c|c|l|}
\hline
no.\!
&
$g$
&
\hfil Basis operators of $A^{\rm max}$
\\
\hline
\multicolumn{3}{|c|}{$n\ne1$}
\\
\hline
1
&
$\forall$
&
$\partial_x$
\\
\hline
2
&
$\lambda T_t(\alpha T+\beta)^\rho$
&
$\partial_x,
\
2n(\alpha T+\beta)T_t^{-1}\partial_t+\alpha n(\rho\!+\!1)x\partial_x+\left(\alpha(\rho\!-\!1)-2nh(t)(\alpha
T+\beta)T_t^{-1}\right)u\partial_u\!\!$
\\
\hline
3
&
$\lambda T_t e^{\alpha T}$
&
$\partial_x,
\
2nT_t^{-1}\partial_t+\alpha nx\partial_x+\left(\alpha-2nh(t)T_t^{-1}\right)u\partial_u$
\\
\hline
4
&
$\lambda T_t$
&
$\partial_x,
\
T_t^{-1}\partial_t-h(t)T_t^{-1}u\partial_u,
\
2nTT_t^{-1}\partial_t+nx\partial_x-\left(2nh(t)TT_t^{-1}+1\right)u\partial_u$
\\
\hline
\multicolumn{3}{|c|}{$n=1$}
\\
\hline
5
&
$\forall$
&
$\partial_x,
\
T\partial_x+T_t\partial_u$
\\
\hline
6
&
$\lambda T_t\left(\frac{\alpha T+\beta}{\gamma T+\delta}\right)^\rho$
&
$\partial_x,
\
T\partial_x+T_t\partial_u$, \\[-2mm]&&
$(\alpha T+\beta)(\gamma T+\delta)T_t^{-1}\partial_t
+\left(\frac12(\rho-1)\Delta+\alpha(\gamma T+\delta)\right)x\partial_x$
\\[-2mm]&&
$\ +\left(\left[-\alpha(\gamma T+\delta)-h(t)(\alpha T+\beta)(\gamma T+\delta)T_t^{-1}
+\frac12(\rho+1)\Delta\right]u+\alpha\gamma T_t x\right)\partial_u$
\\
\hline
7
&
$\lambda T_t e^{\frac{\alpha T+\beta}{\gamma T+\delta}}$
&
$\partial_x,
\
T\partial_x+T_t\partial_u$, \\[-2mm]&&
$\left(\gamma T+\delta\right)^2T_t^{-1}\partial_t+\left(\gamma(\gamma T+\delta)+\frac12\Delta\right)x\partial_x$
\\[-2mm]&&
$\ +\left(\left[-\gamma(\gamma T+\delta)-h(t)(\gamma T+\delta)^2T_t^{-1}
+\frac12\Delta\right]u+\gamma^2T_tx\right)\partial_u$
\\
\hline
8
&
$\!\lambda T_t e^{2\rho\arctan({\alpha T+\beta})}\!\!$
&
$\partial_x,
\
T\partial_x+T_t\partial_u$, \\[-2mm]&&
$\left((\alpha T+\beta)^2+1\right)T_t^{-1}\partial_t
+\alpha\left(\alpha T+\rho+\beta\right)x\partial_x$
\\[-2mm]&&
$\
+\big(\!\left[\alpha(-\alpha T+\rho-\beta)-h(t)\left((\alpha T+\beta)^2+1\right)T_t^{-1}\right]u + \alpha^2 T_tx\big)\partial_u$
\\
\hline
9
&
$\lambda T_t$
&
$\partial_x,
\
T\partial_x+T_t\partial_u,
\
2TT_t^{-1}\partial_t+x\partial_x-\left(2h(t)TT_t^{-1}+1\right)u\partial_u$, \\[-2mm]&&
$T_t^{-1}\partial_t-h(t)T_t^{-1}u\partial_u,
\
T^2T_t^{-1}\partial_t+Tx\partial_x+\left(T_tx-\left(h(t)T^2T_t^{-1}+T\right)u\right)\partial_u$
\\
\hline
\end{tabular}
\end{center}
Here $T=T(t)=\int e^{-n\int h(t)\D t}\D t$,  and
the function $h(t)$ is arbitrary in all cases.
All constants satisfy the same restrictions as in Table~2.

\vspace{-2ex}
\end{table}

\begin{remark}
Using Table~3 it is easy to classify Lie symmetries of equations of the form~\eqref{GBE} with $h=\const$.
For example, if $n\ne1$, then the values of~$g$
that correspond to such equations admitting Lie symmetry extensions are
\[
g=\lambda e^{-nht}(\alpha e^{-nht}+\beta)^\rho,\quad
g=\lambda e^{-nht}e^{\alpha e^{-nht}}\quad\mbox{and}\quad
g=\lambda e^{-nht},
\]
where $\lambda$, $\alpha$ and $\rho$ are nonzero constants,
and the constant $\beta$ is arbitrary (cf.\ Cases~2--4 of Table~3).
The maximal Lie invariance algebras are two-dimensional for the first two values of the parameter-function~$g$.
For the third value of~$g$ the maximal Lie invariance algebra is three-dimensional.
Namely, the equation
$u_t+u^nu_x+hu=\lambda e^{-nht}u_{xx}$, $n\ne0,1$,
admits the maximal Lie invariance algebra $A^{\rm max}=\langle\partial_x, e^{nht}(\partial_t-hu\partial_u),
2\partial_t-nhx\partial_x-hu\partial_u\rangle$.%
\footnote{The third basis element of $A^{\rm max}$ was missed in~\cite{vaga2006a}.}
\end{remark}

\section{Similarity solutions}

One of the most important applications of Lie symmetries is that they provide
a powerful tool for finding closed-form solutions of PDEs.
The Lie reduction method is well known~\cite{Olver1986}.
If a (1+1)-dimensional PDE admits a Lie symmetry generator
of the form $Q=\tau\partial_t+\xi\partial_x+\eta\partial_u$,
then the ansatz reducing this PDE to ODE can be found as a solution of the invariant surface condition $Q[u]:=\tau u_t+\xi u_x-\eta=0$.
In other words, the corresponding characteristic system $\frac{{\rm d}t}{\tau}=\frac{{\rm d}x}{\xi}=\frac{{\rm d}u}{\eta}$
should be solved.
Reductions to algebraic equations can be performed using two-dimensional subalgebras.

\begin{remark}
Some equations from the class~\eqref{GBE} admit discrete point symmetries.
For example, if $g$ and $h$ are odd functions, then the corresponding equation is invariant
with respect to alternating both the signs of $(t,x)$.
We do not use these transformations in the course of the classification of subalgebras
for the corresponding maximal invariance algebras.
\end{remark}

\subsection{Inequivalent subalgebras}\label{SectionISA}

The Lie algebras spanned by the generators presented in Table~1 have the following structure.
In Case~1 the corresponding maximal Lie invariance algebra~$A^{\rm max}$ is one-dimensional (the type~$A_1$).
Here and in what follows we use the notations of~\cite{pate1977a}
for Lie algebras of dimensions less than four.
The algebras in Cases~2, 3 and 5 are two-dimensional.
They are Abelian (the type~$2A_1$) in Case~2 with $\rho=-1$ and in Case~5,
and they are non-Abelian (the type~$A_{2}$) in Case~2 with $\rho\ne-1$ and in Case~3.
The algebras presented in Cases 4 and 6--8 are three-dimensional.
In Case~4 and in Case~6 with $\rho\ne\pm1$ the algebras are of the type $A^a_{3.5}$ with $a=\frac12$
and $a=\big(\frac{\rho-1}{\rho+1}\big)^{\sgn\rho}$, respectively.
If $\rho=\pm1$, then $A^{\rm max}$ from Case~6 is $A_1\oplus A_2$.
In Case~7 $A^{\rm max}$ is of the type $A_{3.2}$,
and in Case~8 it is of the type $A^a_{3.7}$ with $a=|\rho|$.
The five-dimensional algebra presented in Case~9 is $\mathfrak{sl}(2,\mathbb{R})\lsemioplus 2A_1$.

The necessary and highly important step in finding
group-invariant solutions is to construct an optimal system of subalgebras
of the corresponding maximal Lie invariance algebra~$A^{\rm max}$.
This procedure is well described in~\cite{Olver1986}.
In Table~4 we present optimal systems of one- and two-dimensional subalgebras for all cases of Table~1
except Case~9, which corresponds to the classical Burgers equation.%
\footnote{
The optimal systems of one-dimensional subalgebras
for the maximal Lie invariance algebras of generalized Burgers equations
from the subclass~\eqref{GBE_n1_h0} were found earlier in~\cite{doyl1990a} although there is an incorrectness therein.
Namely, the maximal Lie invariance algebras of Case~6 with $\rho=1$ and $\rho=-1$ were supposed to have the same optimal system of subalgebras,
which is not true; cf.\ Table~4.
More specifically, if $\rho=-1$,
then the subalgebra $\langle t\partial_t+(x+at)\partial_x+a\partial_u\rangle$ given in~\cite{doyl1990a} for $\rho=\pm1$
should be replaced by the subalgebra
$\mathfrak g^a_{2.2}=\langle t\partial_t+a\partial_x-u\partial_u\rangle$.
}
(As this equation is linearized by the Hopf--Cole transformation~\cite{cole1951a,hopf1950a} to the linear heat equation,
finding exact solutions of the Burgers equation by Lie reductions is needless.)
These results completely agree with~\cite{pate1977a},
where optimal systems of subalgebras are obtained for all three- and four-dimensional Lie algebras.

\begin{table}[h!]\small \renewcommand{\arraystretch}{1.65}
\begin{center}
\textbf{Table 4.} Optimal systems of one- and two-dimensional subalgebras of algebras given in Table 1.
\\[2ex]
\begin{tabular}
{|l|l|}
\hline
\hfil no.
&
\hfil Subalgebras
\\
\hline
\multicolumn{2}{|c|}{$n\ne1$}
\\
\hline
1
&
$\mathfrak g^{\,}_1=\langle\partial_x\rangle$
\\
\hline
$2_{\rho\ne-1}$
&
$\mathfrak g^{\,}_1=\langle\partial_x\rangle,
\quad
\mathfrak g^{\rho}_{2.1}=\langle t\partial_t+\frac{\rho+1}{2}x\partial_x+\frac{\rho-1}{2n}u\partial_u\rangle,
\quad
\mathfrak g^{\rho}_{2.3}=\langle\partial_x, t\partial_t+\frac{\rho+1}{2}x\partial_x+\frac{\rho-1}{2n}u\partial_u\rangle$
\\
\hline
$2_{\rho=-1}$
&
$\mathfrak g^{\,}_1=\langle\partial_x\rangle,
\quad
\mathfrak g^a_{2.2}=\langle nt\partial_t+a\partial_x-u\partial_u\rangle,
\quad
\mathfrak g^{-1}_{2.3}=\langle\partial_x, nt\partial_t-u\partial_u\rangle$
\\
\hline
3
&
$\mathfrak g^{\,}_1=\langle\partial_x\rangle,
\quad
\mathfrak g^{\,}_{3.1}=\langle2n\partial_t+nx\partial_x+u\partial_u\rangle,
\quad
\mathfrak g^{\,}_{3.2}=\langle\partial_x, 2n\partial_t+nx\partial_x+u\partial_u\rangle$
\\
\hline
4
&
$\mathfrak g^{\,}_1=\langle\partial_x\rangle,
\quad
\mathfrak g^\sigma_{4.1}=\langle\partial_t+\sigma\partial_x\rangle,
\quad
\mathfrak g^{0}_{2.1}=\langle2nt\partial_t+nx\partial_x-u\partial_u\rangle$,
\\
&
$\mathfrak g^{\,}_{4.2}=\langle\partial_t,\,\partial_x\rangle,
\quad
\mathfrak g^{0}_{2.3}=\langle\partial_x,\,2nt\partial_t+nx\partial_x-u\partial_u\rangle,
\quad
\mathfrak g^{\,}_{4.3}=\langle\partial_t,\,2nt\partial_t+nx\partial_x-u\partial_u\rangle$
\\
\hline
\multicolumn{2}{|c|}{$n=1$}
\\
\hline
5
&
$\mathfrak g^{\,}_1=\langle\partial_x\rangle,
\quad
\mathfrak g^a_5=\langle(t+a)\partial_x+\partial_u\rangle,
\quad
\mathfrak g^{\,}_0=\langle\partial_x,\,t\partial_x+\partial_u\rangle$
\\
\hline
$6_{\rho\ne\pm1}$
&
$\mathfrak g^{\,}_1=\langle\partial_x\rangle,
\!\quad
\mathfrak g^{\sigma}_5=\langle(t+\sigma)\partial_x+\partial_u\rangle,
\!\quad
\mathfrak g^{\rho}_{2.1}=\langle t\partial_t+\frac{\rho+1}{2}x\partial_x+\frac{\rho-1}{2}u\partial_u\rangle,
\!\quad
\mathfrak g^{\,}_0=\langle\partial_x,\,t\partial_x+\partial_u\rangle$, \\
&
$\mathfrak g^{\rho}_{2.3}=\langle\partial_x,\,t\partial_t+\frac{\rho+1}{2}x\partial_x+\frac{\rho-1}{2}u\partial_u\rangle,
\quad
\mathfrak g^{\,}_{6.1}=\langle
t\partial_x+\partial_u,\,t\partial_t+\frac{\rho+1}{2}x\partial_x+\frac{\rho-1}{2}u\partial_u\rangle$
\\
\hline
$6_{\rho=-1}$
&
$\mathfrak g^{\,}_1=\langle\partial_x\rangle,
\quad
\mathfrak g^{\sigma}_5=\langle(t+\sigma)\partial_x+\partial_u\rangle,
\quad
\mathfrak g^a_{2.2}=\langle t\partial_t+a\partial_x-u\partial_u\rangle$, \\
&
$\mathfrak g^{\,}_0=\langle\partial_x,\,t\partial_x+\partial_u\rangle,
\quad
\mathfrak g^{-1}_{2.3}=\langle\partial_x,\,t\partial_t-u\partial_u\rangle,
\quad
\mathfrak g^a_{6.4}=\langle t\partial_x+\partial_u,\,t\partial_t+a\partial_x-u\partial_u\rangle$
\\
\hline
7
&
$\mathfrak g^{\,}_1=\langle\partial_x\rangle,
\quad
\mathfrak g^0_5=\langle t\partial_x+\partial_u\rangle,
\quad
\mathfrak g^{\,}_{3.1}=\langle2\partial_t+x\partial_x+u\partial_u\rangle$, \\
&
$\mathfrak g^{\,}_0=\langle\partial_x,\,t\partial_x+\partial_u\rangle,
\quad
\mathfrak g^{\,}_{3.2}=\langle\partial_x,\,2\partial_t+x\partial_x+u\partial_u\rangle$
\\
\hline
8
&
$\mathfrak g^{\,}_1=\langle\partial_x\rangle,
\quad
\mathfrak g^{\,}_8=\langle(t^2+1)\partial_t+(t+\rho)x\partial_x+(x+(\rho-t)u)\partial_u\rangle,
\quad
\mathfrak g^{\,}_0=\langle\partial_x,\,t\partial_x+\partial_u\rangle$
\\
\hline

\end{tabular}
\end{center}
Here $a\in\mathbb R$, $n\ne0$, $\sigma\in\{-1,0,1\}$.
Case~$6_{\rho=1}$ is omitted since it is $\hat G^\sim_{n=1}$-equivalent to~Case~$6_{\rho=-1}$.

\medskip

\small \renewcommand{\arraystretch}{1.75}
\begin{center}
\textbf{Table 5.} Similarity reductions of equations from the class~\eqref{GBE} to ODEs with respect to
\\
one-dimensional invariance algebras given in Table 4.
\\[2ex]
\begin{tabular}
{|c|l|c|c|l|l|}
\hline
\hfil$g(t)$ &\hfil $\mathfrak g$ & $\omega$ &\hfil Ansatz, $u$ &\hfil Reduced ODE
\\
\hline
\multicolumn{5}{|c|}{General value of $n$}
\\
\hline
$\varepsilon t^{\rho}$ & $\mathfrak g^{\rho}_{2.1}$ & $xt^{-\frac{\rho+1}{2}}$ &
$t^{\frac{\rho-1}{2n}}\varphi(\omega)$ & $\varepsilon\varphi''+\left(\frac{\rho+1}{2}\omega-\varphi^n\right)\varphi'+\frac{1-\rho}{2n}\varphi=0$
\\
\hline
$\varepsilon t^{-1}$ & $\mathfrak g^a_{2.2}$ & $x-\frac{a}{n}\ln t$ & $t^{-\frac{1}{n}}\varphi(\omega)$ & $\varepsilon
\varphi''+\left(\frac{a}{n}-\varphi^n\right)\varphi'+\frac1n\varphi=0$
\\
\hline
$\varepsilon e^t$ & $\mathfrak g^{\,}_{3.1}$ & $xe^{-\frac12t}$ & $e^{\frac1{2n}t}\varphi(\omega)$ & $\varepsilon
\varphi''+\left(\frac1{2}{\omega}-\varphi^n\right)\varphi'-\frac1{2n}\varphi=0$
\\
\hline
$1$ & $\mathfrak g^{\sigma}_{4.1}$ & $x-\sigma t$ & $\varphi(\omega)$ & $\varphi''+(\sigma-\varphi^n)\varphi'=0$
\\
\hline
\multicolumn{5}{|c|}{Specific cases for $n=1$}
\\
\hline
$\forall$ & $\mathfrak g^a_5$ & $t$ & $\varphi(\omega)+\dfrac x{t+a}$ & $(\omega+a)\varphi'+\varphi=0$
\\
\hline
$\varepsilon e^{2\rho\arctan t}$ & $\mathfrak g^{\,}_8$ & \rule{0ex}{4ex}$\dfrac{xe^{-\rho\arctan t}}{\sqrt{t^2+1}}$ &
$\dfrac{e^{\rho\arctan t}}{\sqrt{t^2+1}}\varphi(\omega)+\dfrac{xt}{t^2+1}$ & $\varepsilon \varphi''+(\rho\omega-\varphi)\varphi'-\rho
\varphi-\omega=0$
\\[1.5mm]
\hline
\end{tabular}
\end{center}
Here $h(t)=0\bmod\hat G^\sim$, $a\in\mathbb R$, $n\ne0$ and $\varepsilon=\pm1$.
For the algebra $\mathfrak g^{\rho}_{2.1}$ we have $\rho\ne-1$ if $n\ne1$, $\rho\ne\pm1$ if $n=1$,
and $\varepsilon=1$ if $\rho=0$.

\vspace{-4ex}
\end{table}

\subsection{Lie reductions}\label{SectionReduction}

\looseness=-1
{\bf Reductions with one-dimensional subalgebras.}
Ansatzes and reduced equations that are obtained for equations from the class~\eqref{GBE_h0}
by means of one-dimensional subalgebras from Table~4 are collected in Table~5.
We do not consider the reduced equation associated with the subalgebra $\mathfrak g_1=\langle\partial_x\rangle$
because it gives only constant solutions.
We also omit the subalgebra $\mathfrak g_0$ in all cases as it does not satisfy the so-called
transversality condition~\cite{Olver1986} and cannot be utilized in order to construct an ansatz.
Due to several tricks (gauging $h$ to 0 by equivalence transformations,
further simplifying the arbitrary element~$g$ by equivalence transformations
for cases of Lie symmetry extensions
and choosing the simplest representatives among equivalent subalgebras for reductions),
the ansatzes as well as the reduced equations generally have a similar simple structure.

Note that the transformation $y=\varphi^{-n}$ maps the reduced equations from Table~5,
except those related to the subalgebras $\mathfrak g^a_5$ and $\mathfrak g^{\,}_8$,
to so-called Euler--Painlev\'e equations~\cite{sach1988a} of the general form
$yy''+\alpha y'^2+p(\omega)yy'+q(\omega)y^2+\beta y'+\gamma=0$.
Here $p$ and~$q$ are smooth functions of~$\omega$, whereas $\alpha$, $\beta$ and~$\gamma$ are constants.

Only some reduced equations admit order lowering, not to mention the representation of solutions in closed form.
We consider these cases in detail.
This covers and even enhances all existing results on Lie reductions of equations from the class~\eqref{GBE}.%
\footnote{
A reduced equation associated with the subalgebra $\mathfrak g_8$
was obtained in~\cite{wafo2004d} with misprints in signs.
As a result, the expression constructed therein for the unknown function
does not satisfy the corresponding generalized Burgers equation.
}
In what follows $c_0$ and~$c_1$ are arbitrary (integration) constants.

\par\medskip\par\noindent
{\it Subalgebra $\mathfrak g^{\rho}_{2.1}$.}
For $n\ne\pm1$ and $\rho=\frac{1-n}{1+n}$ the corresponding reduced equation is simplified to
$\varepsilon\varphi''-\varphi^n\varphi'+\frac{1}{n+1}(\omega \varphi'+\varphi)=0$
and once integrated, cf.~\cite[Eq.\,(85)]{vaga2006a},
\begin{gather}\label{RE21}
\varepsilon(n+1)\varphi'-\varphi^{n+1}+\omega \varphi+c_0=0.
\end{gather}
It is possible to solve~\eqref{RE21} for the particular value of the integration constant $c_0=0$,
which gives a~one-parametric family of solutions to the equation~\eqref{GBE_h0} with $g(t)=\varepsilon t^{\frac{1-n}{1+n}}$, $\varepsilon=\pm1$,
\begin{gather}
\label{solution21}
u(t,x)=
\frac{t^{-\frac{1}{n+1}}\exp\Big({-\frac{\mu}{n}x^2t^{-\frac{2}{n+1}}}\Big)}
{\left(c_1-2\mu t^{-\frac{1}{n+1}}\int\exp\Big(-\mu x^2t^{-\frac{2}{n+1}}\Big)\D x\right)^\frac1n},
\qquad
\mu={\frac{n}{2\varepsilon(n+1)}}.
\end{gather}

If $\mu>0$, then these solutions can be represented in terms of the error
function $\erf$, $\erf(\theta)=\frac{2}{\sqrt{\pi}}\int_0^{\theta}\!e^{-s^2}\!\D s$, as
\begin{gather}
\label{solution21erf}
u(t,x)=
\frac{t^{-\frac{1}{n+1}}\exp\Big({-\frac{\mu}{n}x^2t^{-\frac{2}{n+1}}}\Big)}
{\left(c_1-\sqrt{\pi\mu}\,\erf\!\Big(\sqrt{\mu}\,xt^{-\frac{1}{n+1}}\Big)\right)^\frac1n}.
\end{gather}

\par\medskip\par\noindent
{\it Subalgebras $\mathfrak g^a_{2.2}$.}
In the case $n=1$ the subalgebra of this family with $a=0$ gives a reduced equation, which can be once integrated
to $\varphi'+\ln(\varphi'-1)=\frac{\varphi^2}{2\varepsilon}+c_0$, or $(\varphi'-1)e^{\varphi'-1}=\tilde c_0e^{\frac{\varphi^2}{2\varepsilon}}$.
The last equation can be solved with respect to~$\varphi'$
in terms of the Lambert function $W(y)$, which is implicitly defined by the equation $W e^W=y$, as
$\varphi'=1+W\big(\tilde c_0e^{\frac{\varphi^2}{2\varepsilon}}\big)$.
Thus, the general solution of the reduced equation under consideration can be implicitly represented
via a single quadrature involving the Lambert function.
It is also possible to express $\varphi$ via~$\varphi'$ from the once integrated equation
and then use the fact that the general solution of an equation $\varphi=F(\varphi')$
is represented in the parametric form as $\omega+c_1=\int \zeta^{-1}F'(\zeta)\,{\rm d}\zeta$, $\varphi=F(\zeta)$.

\par\medskip\par\noindent
{\it Subalgebra $\mathfrak g^{\,}_{3.1}$.}
The corresponding reduced equation in the case $n=-1$ can be once integrated to
$\varepsilon \varphi '+\frac12\omega\varphi -\ln \varphi+c_0=0$, but nothing more.%
\footnote{
The reduced equations associated with the subalgebras~$\mathfrak g^{\rho}_{2.1}$ and~$\mathfrak g^{\,}_{3.1}$
were numerically solved in~\cite{sach1988a}.
In fact, an equivalent reduction to the first one was considered therein for the value $g=(t+1)^\rho$,
which is equivalent to $g=t^\rho$ up to shifts of $t$.
}

\par\medskip\par\noindent
{\it Subalgebras $\mathfrak g^\sigma_{4.1}$.}
For convenience, we extend $\sigma$ to an arbitrary~$a\in\mathbb R$ from the very beginning.
The ansatz $u=\varphi(x-at)$ obtained for $g(t)=1$ (for $n=1$ it corresponds to the classical Burgers equation)
leads to the reduced equation $\varphi''+(a-\varphi^n)\varphi'=0$, which can be
once integrated,
\begin{gather*}
\varphi'-\frac{1}{n+1}\varphi^{n+1}+a\varphi+c_0=0, \quad\text{if}\quad n\ne-1,
\\
\varphi'+a\varphi-\ln \varphi+c_0=0, \quad\text{if}\quad n=-1.
\end{gather*}
After one more integration, the general solutions of these equations are implicitly expressed via single quadratures.
For some values of constant parameters it is possible to find explicit formulae for solutions.
An obvious case $n=1$, which corresponds to the classical Burgers equation, is not interesting.
Another case with an explicit solution is $c_0=0$, which results, depending on values of $a$, in
two different families of solutions to~\eqref{GBE_h0} with $g(t)=1$,
\begin{gather}
u(t,x)=\left({\frac{a(n+1)}{1+c_1e^{an(x-at)}}}\right)^{\frac1n}
\quad \text{if} \quad a\ne0, \nonumber
\\
\label{solution41a}
u(t,x)=\left({\frac{n+1}{c_1-nx}}\right)^{\frac1n}
\quad \text{if} \quad a=0.
\end{gather}

\par\medskip\par\noindent
{\it Subalgebras $\mathfrak g^a_5$.} The corresponding reduced equation is
$(\omega+a)\varphi'+\varphi=0$.
It gives a degenerate solution of~\eqref{GBE_n1_h0} for an arbitrary value of $g(t)$,
\begin{gather}
\label{solution5}
u(t,x)=\frac{x+c_0}{t+a}.
\end{gather}

\medskip\par\noindent
\looseness=-1
{\bf Reductions with two-dimensional subalgebras.}
Such reductions, which gives algebraic reduced equations,
do not lead to nontrivial exact solutions for equations from the class~\eqref{GBE}.
Thus, the subalgebra~$\mathfrak g^{\,}_{0}$ does not satisfy the transversality condition
and hence it is not associated with a Lie ansatz for~$u$.
The reductions with respect to~$\mathfrak g^{\rho}_{2.3}$ and~$\mathfrak g^{\,}_{3.2}$
give only the trivial (identically zero) solution.
A solution is invariant with respect to~$\mathfrak g^{\,}_{4.2}$ if and only if it is a constant.
Using the subalgebra~$\mathfrak g^{\,}_{4.3}$ we obtain the solution~\eqref{solution41a} with $c_1=0$.
The only solution that is invariant with respect to the subalgebra~$\mathfrak g^{\,}_{6.1}$ has the form~\eqref{solution5} with $a=c_0=0$.
The reduction using the subalgebras~$\mathfrak g^a_{6.4}$ results in a solution only if $a=0$,
and this solution has the form~\eqref{solution5} with $a=0$.

\subsection{Generation of exact solutions for equations from the initial class}\label{SectionGeneration}

Using solutions of~\eqref{GBE_h0} and~\eqref{GBE_n1_h0} obtained in Section~\ref{SectionReduction} and the transformation~\eqref{tr_tau},
one can derive solutions of equations from the initial class~\eqref{GBE} with arbitrary values of~$h(t)$,
\begin{gather*}
\arraycolsep=0ex
\begin{array}{rll}
\text{\,({\it i})}\,\,
&
u_t+u^nu_x+h(t)u=\varepsilon T^{\frac{1-n}{1+n}}e^{-n\int h(t)\D t}u_{xx}\colon
\quad
&
u=\dfrac{T^{-\frac{1}{n+1}}\exp\big({-\frac{\mu}{n}x^2T^{-\frac{2}{n+1}}}\big)e^{-\int h(t)\D t}}
{\big(c_1-2\mu T^{-\frac{1}{n+1}}\int e^{-\mu x^2T^{-\frac{2}{n+1}}}\D x \big)^\frac1n},
\\[2ex]
\text{\,({\it ii})}\,\,
&
u_t+u^nu_x+h(t)u=e^{\,-n\!\int h(t)\D t}u_{xx}\colon
&
u=\left({\dfrac{a(n+1)}{1+c_1e^{an(x-aT)}}}\right)^{\frac1n}\!e^{-\int h(t)\D t},
\\[1ex]
&
\phantom{u_t+u^nu_x+h(t)u=e^{n\int h(t)\D t}u_{xx}:}
&
u=\left({\dfrac{n+1}{c_1-nx}}\right)^{\frac1n}\!e^{-\int h(t)\D t},
\\[2.5ex]
\text{\,({\it iii})}\,\,&
u_t+uu_x+h(t)u=g(t)u_{xx}\colon
&
u=\dfrac{x+c_0}{\int e^{-\int h(t)\D t}\D t+a}\,e^{-\int h(t)\D t}.
\end{array}
\end{gather*}
Here $a$, $c_0$ and $c_1$ are arbitrary constants, $\varepsilon=\pm1$,
$\mu={\frac{n}{2\varepsilon(n+1)}}$, and the function~$T=T(t)$ is defined as
\begin{gather*}
T(t)=\int e^{-n\int h(t)\D t}\D t.
\end{gather*}
Before the application of the transformation~\eqref{tr_tau},
the solution~\eqref{solution21} can be additionally extended
by transformations from the equivalence group~$\hat G^\sim_{h=0}$.
(Shifts of~$x$ belong to the intersection of the maximal Lie symmetry groups of all equations
from the class~\eqref{GBE} (resp.\ from the class~\eqref{GBE_h0}) and just transform solutions,
whereas shifts of~$t$ and scalings are rather equivalence transformations and change also
the arbitrary element~$g$).

The equation ({\it ii}) can be rewritten as
\begin{equation}\label{Eq_Burgers_reducibletoconstcoeff}
u_t+u^nu_x-\frac1n\frac{k_t}ku=ku_{xx},
\end{equation}
where the functions $k=k(t)$ and $h=h(t)$ are related via the formula $k=e^{\,-n\!\int h(t)\D t}$.
For $n=1$ the equation \eqref{Eq_Burgers_reducibletoconstcoeff} coincides with equation (3.262) in~\cite[p.~90]{sach2009a},
which includes the Burgers model for turbulence but with variable diffusivity (applicable to modeling of acoustic waves in the atmosphere).
We have constructed two families of exact solutions for the equation ({\it ii}).
The behavior of the solution
\begin{equation}\label{Eq_SolforGRaphs}
u=\left({\dfrac{a(n+1)}{1+c_1e^{an(x-aT)}}}\right)^{\frac1n}\!e^{-\int h(t)\D t}
\end{equation}
for two values of~$n$ and two forms of the variable diffusivity coefficient~$h$ is illustrated at Fig.~1.

\begin{figure}
\includegraphics{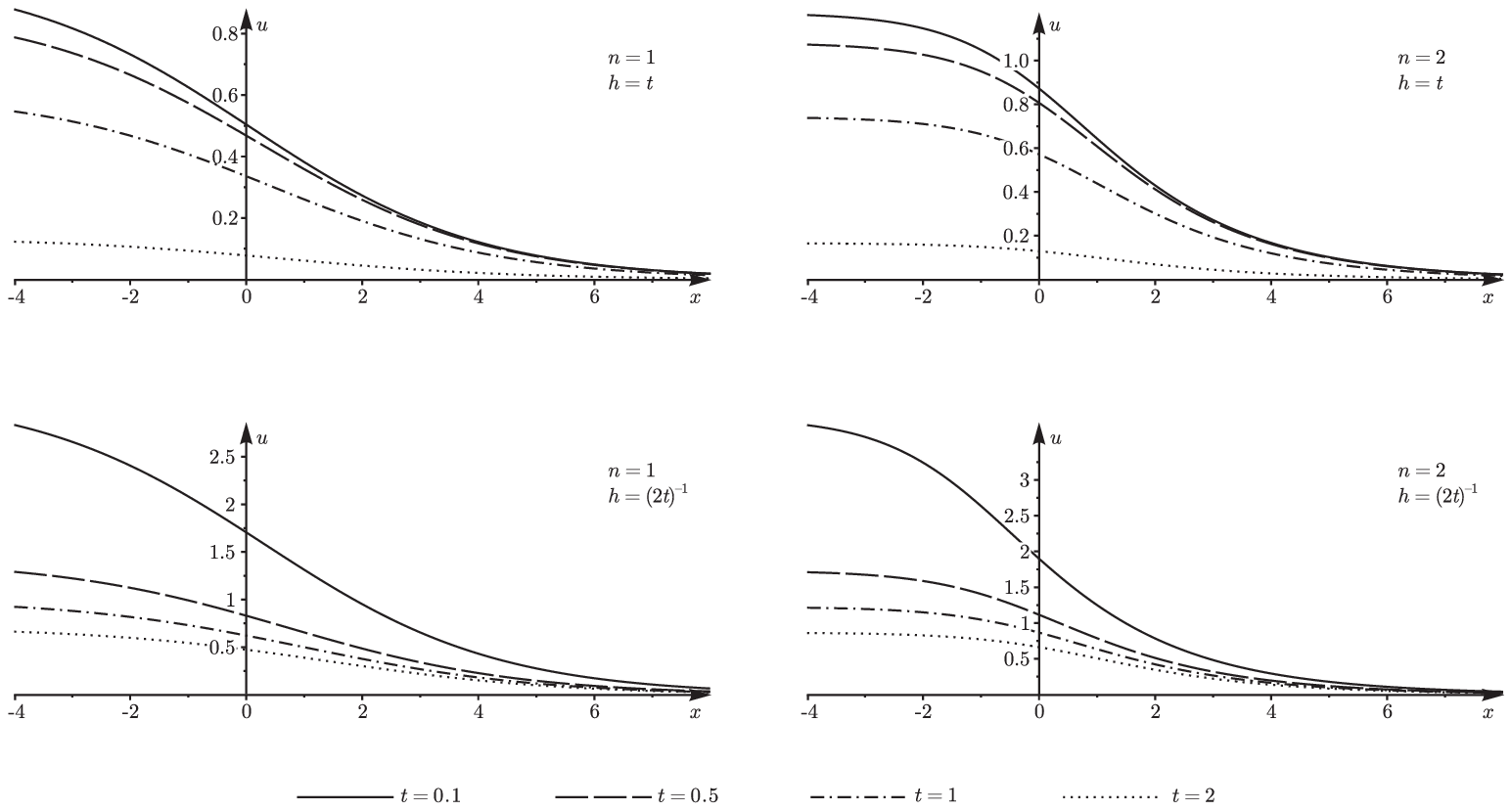}
\vspace{3mm}
\caption{\footnotesize The behavior of the solution~\eqref{Eq_SolforGRaphs} for $c_1=1$, $a=0.5$ and values of $n$ and~$h$ indicated on the graphs.
\noprint{Here
$n=1$, and $h=t$ at Fig.~a,
$n=2$, and $h=t$ at Fig.~b;
$n=1$, and $h=(2t)^{-1}$ at Fig.~c,
$n=2$, and $h=(2t)^{-1}$ at Fig.~d.
}
}
\end{figure}

\section{Generalized Burgers equations linearizable\\ to the heat equation}\label{Section_LGBE}

The remarkable property of the classical Burgers equation
\begin{gather}\label{BE}
\hat u_{\hat t}+\hat u\hat u_{\hat x}=\lambda \hat u_{\hat x\hat x}
\end{gather}
is that it is linearizable to the heat equation
\begin{gather}\label{HeatE}
\hat v_{\hat t}=\lambda \hat v_{\hat x \hat x}
\end{gather}
by the Hopf--Cole transformation $\hat u=-2\lambda\hat v_{\hat x}/\hat v$.
More precisely, the Hopf--Cole transformation reduces the equation~\eqref{BE} to the equation
$\left(\hat v\partial_{\hat x}-\hat v_{\hat x}\right)[\hat v_{\hat t}-\lambda \hat v_{\hat x \hat x}]=0$,
which can be integrated once with respect to~$x$,
and the ``integration constant'', which is an arbitrary function of~$t$,
can be neglected due to the freedom in choosing the function~$\hat v$.
Thus exact solutions for~\eqref{BE} can be easily obtained from exact solutions of the linear heat equation~\eqref{HeatE}.

In~\cite{vaga2011a} variable-coefficient generalized Burgers equations of the form
\begin{gather}\label{Lin1}
u_t+uu_x+h u=\lambda e^{-ht}u_{xx}
\end{gather}
with $\lambda=1/2$ were linearized in different ways to linear parabolic equations
that are similar to the classical heat equation with respect to point transformations.
In our opinion, the direct linearization of~\eqref{Lin1} to the heat equation is more efficient.
It can be done using the composition of an equivalence point transformation of the class~\eqref{GBE}
that maps~\eqref{Lin1} to~\eqref{BE} and the Hopf--Cole transformation.
Namely, the equation~\eqref{Lin1} is linearized to~\eqref{HeatE} by the transformation
\begin{gather*}
\hat t=-\frac 1{h}{e^{-h t}},
\quad
\hat x=x,
\quad
-2\lambda\frac{\hat v_{\hat x}}{\hat v}=e^{h t}u.
\end{gather*}

More generally, all equations from the class~\eqref{GBE} that are linearized to~\eqref{HeatE} have the form
\begin{gather}\label{GBE_Lin}
u_t+uu_x+h(t)u=\lambda e^{-\int h(t)\D t}u_{xx}.
\end{gather}
The corresponding transformation
\begin{gather*}
\hat t=\int e^{-\int h(t)\D t}\D t,\quad \hat x=x,\quad -2\lambda\frac{\hat v_{\hat x}}{\hat v}=e^{\int\! h(t)\D t}u
\end{gather*}
gives the formula for generating solutions of the equation~\eqref{GBE_Lin} from solutions of the heat equation
\begin{gather*}
u(t,x)=-2\lambda e^{-\int h(t)\D t}\frac{\hat v_x(\hat t,x)}{\hat v(\hat t,x)},
\quad
\mbox{where}
\quad
\hat t=\int e^{-\int h(t)\D t}\D t.
\end{gather*}

Consider, for example, the exact solution $\hat v=ce^{a\hat x +a^2\lambda\hat t}(\hat x+2a\lambda\hat t)$ of the heat equation~\eqref{HeatE}.
Here $c$ and $a$ are arbitrary nonzero constants.
Using the last transformation, we get an exact solution of~\eqref{GBE_Lin},
\begin{gather*}
u(t,x)=-2\lambda\frac{ax+2a^2\lambda \int e^{-\int h(t)\D t}\D t+1}{x+2a\lambda\int e^{-\int h(t)\D t}\D t} e^{-\int h(t)\D t}.
\end{gather*}

\section{Conclusion}

In this paper we have carried out the exhaustive Lie symmetry analysis of equations from the class~\eqref{GBE}.
The consideration is essentially based on the complete description of
the equivalence groupoids of the entire class and certain its subclasses,
which is presented in Theorems~\ref{Theorem_GenBurgers_G}--\ref{Theorem_GenBurgers_EquivGroupoid}
and their corollaries and involves the notion of normalized class of differential equations.
One more essential point is that for the subclass~\eqref{GBE_n1} singled out by the constraint $n=1$
it is necessary to find its generalized extended equivalence group,
in which the transformation components for~$x$ and~$u$
nonlocally depend on the arbitrary element~$h$.
The equivalence generated by the usual equivalence group of the class~\eqref{GBE} or the above subclass
is too weak for an effective use.
In particular, transformations from the generalized extended equivalence group of the subclass~\eqref{GBE_n1}
play an important role in the linearization of equations from this subclass.

The group classification problem for the entire class~\eqref{GBE} up to general point equivalence
has been solved by the reduction to the same problem for the subclass~\eqref{GBE_h0}
due to the fact that the arbitrary element~$h(t)$ is gauged to zero using a parameterized family of equivalence transformations.
The corresponding classification list is presented in Table~1,
whereas Table~3 contains the complete list of Lie symmetry extensions for the class~\eqref{GBE}
without simplifying the forms of the arbitrary elements by equivalence transformations.

The results on Lie symmetries have been applied to finding exact solutions of equations from the class~\eqref{GBE}.
All possible inequivalent Lie reductions have been carried out in a systematic way.
Both general point equivalence of equations from the class~\eqref{GBE} and equivalence of subalgebras
with respect to internal automorphisms of the corresponding maximal Lie invariance algebras have been involved in the consideration,
which significantly reduced the number of different Lie reductions to be studied
and also simplified the corresponding ansatzes and reduced equations.
In spite of the concise presentation, families of exact solutions
constructed in the paper to equations from the class~\eqref{GBE}
include, but are not limited by, all closed-form solutions presented for these equations in the literature.

\bigskip\par\noindent{\bf Acknowledgements.}
The authors are grateful to Prof. C. Sophocleous for pointing out many relevant bibliographic references
and to Prof. V.M. Boyko for useful discussions.
The research of ROP was supported by the Austrian Science Fund (FWF), project P25064.

\end{document}